\documentclass{article}
\RequirePackage{fix-cm}

\usepackage[utf8]{inputenc}
\usepackage{graphicx}
\usepackage{amssymb}
\usepackage{latexsym}
\usepackage{algorithm}
\usepackage{algorithmic}
\usepackage{colortbl}
\usepackage{url}
\usepackage[english]{babel}
\usepackage{microtype} 
\usepackage{tabularx} 
\usepackage{booktabs} 
\usepackage{graphicx}
\usepackage{listings}
\usepackage{geometry}
\usepackage{color}
\usepackage{textcomp}
\usepackage{xcolor}
\usepackage{natbib}

\colorlet{light-gray}{gray!20}

\definecolor{listinggray}{gray}{0.9}
\definecolor{lbcolor}{rgb}{0.9,0.9,0.9}
\begin{document}

\title{Should Decorators Preserve the Component Interface?}

	\author{V. Niculescu,
	A. Sterca,
	D. Bufnea \\
	Faculty of Mathematics and Computer Science, \\
	Babe\c s-Bolyai University, Romania \\
	}

\date{}

\maketitle


\begin{abstract}
{\em Decorator} design pattern is a well known pattern that allows dynamical attachment of additional functionality to an object. Decorators have been proposed as flexible alternative to subclassing for extending functionality.
Still, the {\em Decorator} pattern has certain limitations, especially related to the fact that in its classical form it is constrained to a single interface, which is implicitly defined  by the type of the concrete components that we intend to decorate. 
Another problem associated to the {\em Decorator} pattern is related to the linear composition of the decorations, which  could lead to problems in accessing the newly added responsibilities. \\
In this context, the paper presents
 variants of the {\em Decorator} pattern: {\em MixDecorator} and {\em D2Decorator}, and a variant specific only to C++ language based on templates -- {\em HybridDecorator}. 
 {\em MixDecorator} could be considered a new enhanced version of the {\em Decorator} pattern  that eliminates some constraints of the {\em Decorator} pattern, but also it could be used as a base of a general extension mechanism. The main advantage of using {\em MixDecorator} is that it allows  direct access to all newly added responsibilities, and so, we may combine different interface-responsibilities (newly added public methods) and operate with them directly and in any order, hiding the linear composition of the decorations.   
 {\em D2Decorator}  is a variant based on a double-dispatch mechanism, which is used for connecting the actual decorator that has to receive a certain message call.  
 The {\sf C++} metaprogramming mechanism based on templates allows an interesting hybrid variant of the {\em Decorator} -- {\em HybridDecorator}, which mixes on-demand defined inheritance with composition.\\
 Using these variants of the {\em Decorator} pattern we are not longer limited to one single interface; the set of the messages that could be sent to an object could be enlarged, and so, we may consider that using them,  we can dynamically change the type of objects.

\end{abstract}



\section{Introduction}
\label{sec-introduction}

The authors of {\it Gang of Four} Design Patterns book \cite{gof} argue in favour of {\it object composition} over {\it class inheritance}, and
{\it Decorator} pattern is one of the patterns that well expresses this issue.
The classical {\em Decorator} pattern offers extensions of  objects functionality in order to modify their behaviour.
The properties of the designs based on {\em Decorator} are similar to those  of the corresponding designs based on inheritance, but the variations are in this case even more modularized; we obtain fine-grained modularization.  
	{\em Decorator}-based designs define variations that are reusable with any class in the basic component hierarchy, and also the variations could be applied in different combinations and dynamically.

The usual agreement is  that decorators and the original component class  share a common set of operations (interface), 
 but this requirement is mostly due to the classical solution proposed for the problem. 
 Transparent enclosure condition imposes only the fact that a decorated component  should be able to receive any message from the component interface.

 %

It  may be considered that each method that belongs to an object interface (i.e. the set of all methods that could be invoked for that object) corresponds to a responsibility of that object, and we will refer to these methods that characterize an object behaviour as {\it interface-responsibilities}.
We argue that the  {\em Decorator} pattern should allow defining decorations that add new interface-responsibilities, not just changing the behaviour of an existing one. This would be natural if we analyse the pattern intent and applicability.


In order to overcome this limitation to one interface,
 the paper introduces and discusses  new enhanced {\em Decorator} variants: {\em MixDecorator}, {\em D2Decorator}, and a C++ meta-programming approach based on templates.

{\it MixDecorator} is an enhanced version of {\em Decorator} pattern that does not just eliminate some constraints of the classical pattern (e.g. limitation to one interface), but it introduces significant flexibility and abstraction,  allowing the definition of a general extension mechanism.
It relies on  recursive dispatcher methods  for finding the invoked methods.

A double-dispatch mechanism could be also used in order to pass over the linear composition constraints; this led to another variant which we choose to call {\em D2Decorator}.  This variant imposes building a new class (a dispatcher) for each newly added  interface-responsibility, but  allows dynamic extensibility in a structured way. 

{\sf C++} templates represents a powerful static mechanism, that allows behaviour infusion or dependency injection -- as in the case of {\sf C++}  policies. They can be used for creating a hybrid {\em Decorator} variant that combines static and dynamic applications.

The paper is structured as follows:
next section succinctly describes the classical version of the {\em Decorator} pattern and emphasizes the constraints  and limitations imposed by it. Section \ref{sec:mixdecorator} describes the general definition  of the {\em MixDecorator} pattern.  At the implementation phase the pattern could be simplified, but there are different simplifications that  are  language dependent. Section \ref{sec:implementation} presents these possibilities for {\sf Java} and {\sf C\#},  {\sf C++}, and some details about the {\sf Python} implementation.
The {\em D2Decorator} is presented in Section \ref{sec:double-dispatch} with its advantages and disadvantages, and 
the {\em Hybrid Decorator} is discussed in section \ref{sec:hybrid}.
The analysis of these variants and the conclusions are presented in section \ref{sec:conclusions}.

\section{Decorator pattern}
\label{sec:decorator}

\begin{figure}
\centering{
\includegraphics[scale=0.45]{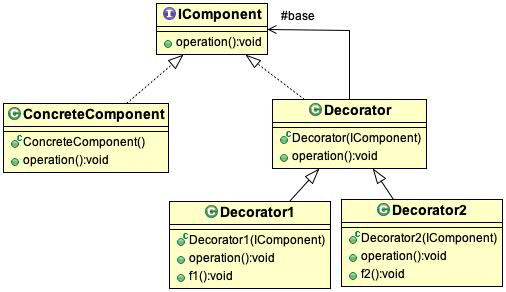}
}
\caption{The class diagram of the standard {\em Decorator} pattern.}
\label{fig:basic}
\end{figure}

The {\em Decorator} pattern is a structural design pattern used to extend or alter the functionality of objects  by wrapping them into  instances of selected decorator classes.  The variation of the functionality is very well modularized -- only one class is defined per each variation (\cite{gof,ST04}).  Essentially, this allows decoration based on linear composition of some independent decorations.

The {\em intent} of this pattern is to
attach additional responsibilities to an object dynamically, and to provide a flexible alternative to subclassing for extending functionality;  its  {\em applicability} is:\vspace{-0.2cm}
\begin{itemize}
\item to add responsibilities to individual objects dynamically and transparently, that is, without affecting other objects;
\item for responsibilities that can be withdrawn;
\item when extension by subclassing is impractical because:
\begin{itemize}
	 \item 
 an explosion of subclasses is needed to support every combination, or
\item a class definition may be hidden or otherwise unavailable for subclassing.
\end{itemize}
\end{itemize}

All these do not suggest that we need to impose that the decorators should conform to the interface of the component it decorates. This is only imposed by the proposed solution. 

 Figure \ref{fig:basic} shows the corresponding class diagram of the classical solution.

An instance of the {\tt\small ConcreteComponent} class could be 'decorated' with an instance of {\tt\small Decorator1}, or {\tt\small Decorator2}, and the result could be decorated again with another decorator instance.
A forwarding semantic is associated to decorators  (\cite{generic_wrappers}) that could redefine the base method {\tt\small operation}, but
each redefinition of the method {\tt\small operation} inside the decorators should invoke the {\tt\small operation} on the aggregated object ({\tt\small base}).
As the component class and the class of each decorator share the same base class, multiple decorators can be applied to the object in order to incrementally modify behaviour.
This means that the modifications could be done also at the run-time not only at the design time. This allows changes to be applied to objects in response to specific conditions such as user-selected options or business rules. 

The solution of the pattern is based on a combination between inheritance and composition: {\tt\small Decorated} is derived from {\tt\small IComponent}  but in the same time wraps an {\tt\small IComponent}. So, theoretically, the associated semantic would be that a {\tt\small Decorator} ``is-a'' but also ``has-a'' {\tt\small IComponent}; still in this case composition and inheritance are used only as implementation mechanisms.



\subsection{Limitations of the Classical {\em Decorator} Pattern}


A {\em Decorator} based design could encounter some problems such as:  
\begin{itemize}
    \item 
lack of object identity  (a decorator and its component are not identical);
   \item no late binding -- since
a decorator and its component interact via forward semantics, which does not ensure late binding (\cite{generic_wrappers}); 
   \item fragile base-class problem -- when the component interface is changed
(\cite{Kniesel2004EvolvablePI,Sabane2016FragileBP,effective_java}); and 
   \item the
limitation of the decorators to the component interface.

\end{itemize}

We will discuss here in more details  the last limitation and how it could be overcome.

As a possible usage scenario, we may consider that we have $n$ new interface-responsibilities intended to be defined as decorations for a base class of {\tt\small IComponent} type. These responsibilities are defined as methods -- {\tt\small f1, f2, \dots, f{\it n}}.
As the pattern specifies, $n$ decorator classes will be defined ({\tt\small Decorator1, Decorator2 $\dots$ Decorator{\it n}}), each defining the corresponding method, and they are all derived from the decoration class {\tt\small Decorator}.
Theoretically, we may obtain any combination of decorations, but we only have  the base class interface available ({\tt\small ICompoment}).

So, if there are some responsibilities that are really new interface-responsibilities (that change the object interface) and they are not used just to alter the behaviour of the operations defined in the base class, they will be accessible only if the last added decoration is the one that defines them. 
More specifically, if the responsibility {\tt\small f1} is a new interface-responsibility and it is defined in the class {\tt\small Decorator1}, then the corresponding message could only be sent  to an object that has the {\tt\small Decorator1} decoration, and  if it is used through a reference of {\tt\small Decorator1} type. The following Java code snippet emphasizes this situation (Listing \ref{lst:limitation}).

	\begin{figure}
		\centering{
		\begin{lstlisting}[caption = {The  access to a decorated functionality defined as a new interface-responsability.},captionpos=b, label =lst:limitation, frame=single, backgroundcolor=\color{light-gray}, basicstyle=\small\ttfamily, language=Java, numbers=left, numberstyle=\tiny             
		\color{black}]
		IComponent o = new Decorator1(new Decorator2(new ConcreteComponent())));
		((Decorator1)o).f1();
		IComponent oo = new Decorator2(new Decorator1(new ConcreteComponent())));
		// ((Decorator1).oo).f1(); ERROR
		((Decorator1)oo.getBase()).f1(); //an improper solution
		\end{lstlisting}
		}
	\end{figure}

The code in Listing \ref{lst:limitation} suggests a possible solution, but this is an improper solution since it has obviously several drawbacks:\vspace{-0.2cm}
\begin{itemize} 
\item we have to invoke  an additional operation that allows decoration removal -- {\tt\small getBase()}; in case we don't have it, there is no solution;
\item if {\tt\small Decorator2} is removed, its  added functionality is lost;
\item if there are several decorations that should be removed then several additional operations are necessary, and all corresponding added behaviour is lost;
\item if we don't know  the exact position (order) of the searched decoration, the code becomes very complex (some sort of reflection has to be used).
\end{itemize}

In fact, removing decorations in order to reveal interface-responsibilities is not a real solution since it breaks the way in which decorated objects are supposed to be used. Also, it is an ad-hoc workaround that is based on knowing the order in which decorations were added.

\subsection{New Forces for Decorator}

Based on the previous analysis we enlarge the {\em Decorator} pattern {\em 'forces'}  to overcome  the analysed limitation:

\begin{enumerate}
	\item Adding new capabilities (including interface-responsibilities) should be possible to clients. 
	\item The different capabilities should be decoupled and reusable.
	\item  Easy to change, e.g. withdraw or add capabilities. 
	\item All newly added responsibilities should be directly accessible to the client.
	\item Assure good efficiency and  extendability.
\end{enumerate}

\section{Applications and examples}
\label{sec:applications}

Many  situations where decorations imply adding new interface-responsibilities could be encountered. 
The applications that were initially designed based on the classical {\em Decorator}, and which define new interface-responsibilities for the decorated objects need to be more carefully treated and adjusted. 

Two examples are discussed next.

\begin{figure}
	\centering{
		\includegraphics[scale=0.4]{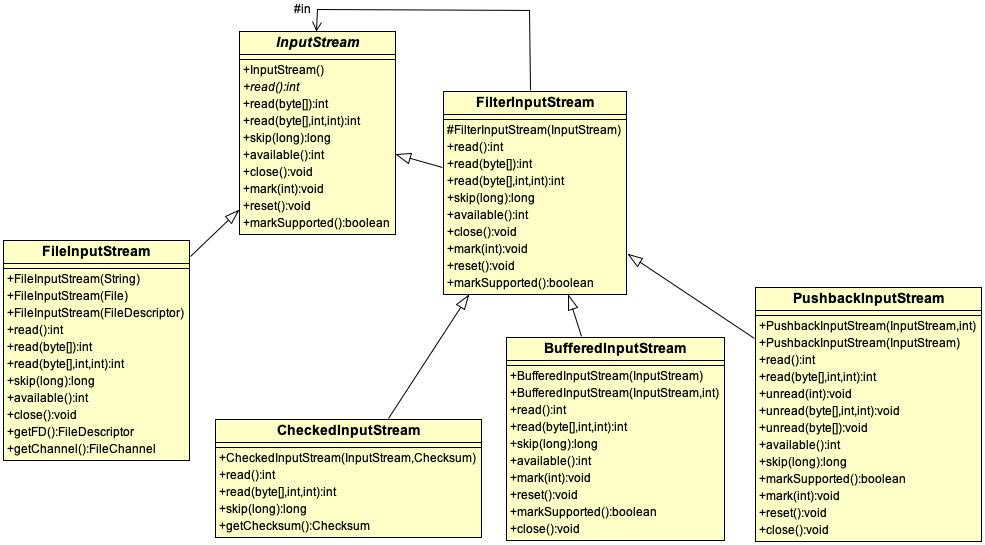}
	}
	\caption{Part of the class diagram of Java {\tt\small InputStream} hierarchy.}
	\label{fig:iostream}
\end{figure}

{\bf Example [Java IO streams].} 
The definition of Java IO streams is a classical example of {\em Decorator} usage.
We may consider the {\tt\small InputStream} hierarchy  from Java IO streams package -- Figure \ref{fig:iostream}. In this case, {\tt\small FilterInputStream} corresponds to the {\tt\small Decorator}, and it is derived from {\tt\small InputStream} that corresponds to {\tt\small IComponent}. As it can be noticed {\tt\small FilterInputStream} preserves the interface of {\tt\small InputStream}.
There are several decoration classes derived from {\tt\small FilterInputStream} such as {\tt\small PushBackInputStream} that defines three {\tt\small unread} methods, which are not defined in the {\tt\small FilterInputStream} interface; {\tt\small BufferedInputStream} that just alters the behaviour of the standard {\tt\small InputStream} interface; or {\tt\small CheckedInputStream} that  maintains a checksum of the data being read and allows using it using the method {\tt\small getChecksum}.

In a practical usage, we may combine them and decorate a certain stream --
e.g. {\tt\small FileInputStream}, first with the {\tt\small PushBackInputStream} decorator, and then with
the {\tt\small BufferedInputStream} or/and with\\ the {\tt\small CheckedInputStream}.
If we would like to use the {\tt\small unread()}  method, this is not longer directly available.

\begin{lstlisting}[caption =  {Java {\tt\small InputStream} example}, captionpos=b, label =lst:iostream, frame=single, backgroundcolor=\color{light-gray}, basicstyle=\small\ttfamily, language=Java, numbers=left, numberstyle=\tiny             
\color{black}]
InputStream pi = new BufferedInputStream(new CheckedInputStream(
   new PushbackInputStream(new FileInputStream("input")), new CRC32()));
//	pi.unread(); ERROR
\end{lstlisting}

Since, the class {\tt\small FilterInputStream} does not provide an operation as {\tt\small getBase()}, not even the simplistic solution presented before in the previous section is  possible. (The class {\tt\small FilterInputStream} has a field {\tt\small in} but it is protected and so inaccessible; a solution could be to specialize all the classes derived from {\tt\small FilterInputStream} and define for them a {\tt\small  getBase()} method that returns {\tt\small in}, but this is obviously improper.)
 If a method of type {\tt\small getBase()} would be provided for {\tt\small FilterInputStream}, then one of the enhanced variants of the {\em Decorator} that we propose here could be used, and so we could eliminate the constraints of the current implementation.

{\bf Example [ReaderDecorator].}
We consider an application that defines  {\em text analyzer}  decorations which decorate a {\tt\small Reader} (an object that could retrieve from a stream, a single character, an entire line, or a specified number of characters). There are examples of such readers in Java and also in C\# - (Java: the {\tt\small Reader} class;  C\# - the {\tt\small  StreamReader} class). 
In order to offer a {\em Decorator} infrastructure for such a class
 we need to
define a {\tt\small ReaderDecorator}  class (that corresponds to {\tt\small Decorator}) that wraps inside a {\tt\small Reader} object.

We may first define a decoration that is able to count the number of already read characters, and next we may add another decoration oriented to words that provides methods such as {\tt\small readWord()} (able to read the next word), and {\tt\small  getNoWords()} (that returns the number of words already read).  
For these, there are  new interface-responsibilities that should be used:
{\tt\small getNoChars(),  readWord(), getNoWords()}.
These new responsabilities imply also overwriting the {\tt\small readChar()} method in order to update the {\tt\small no\_of\_chars} and {\tt\small no\_of\_words} attributes defined by the new decorators.

In a further step of the development, a sentence oriented decorator could be added; it defines methods such as {\tt\small readSentence()}  and {\tt\small getNoSentences()}. They also imply a new attribute definition {\tt\small no\_of\_sentences} that should be updated by {\tt\small readChar()}.

The use of decorations is appropriate since if we do not  need to read or count sentences we don't have  to add the {\tt\small SentenceDecorator}, and similarly for {\tt\small WordDecorator}, or {\tt\small CharCounterDecorator}.
The three decorators could be added in any order, and also could be retrieved if they are no longer necessary (to assure efficiency). 

Further developments are possible, by adding  other different kinds of decorators oriented on text reading. Examples could be  decorators able to read a special kind of text files (xml, html, or others).

\section{{\em MixDecorator}}
\label{sec:mixdecorator}
In order to overcome the unique interface limitation of the {\em Decorator} pattern, an enhanced variant named {\em MixDecorator} is proposed. Its first proposal has been done by \cite{Nic15}, and here is presented an adapted improved version of that.





\subsection{The {\em MixDecorator} Solution:}
\label{sec:def-mixdecorator}
The structure of the {\em MixDecorator} is inspired by the {\em Decorator} pattern and it is very similar to it, but there are several important differences that allow  achieving the enhanced {\em 'forces'.}



The structure of this solution is presented in Figure \ref{fig:mixdecorator}. 
The {\tt\small Decorator} class has almost  the same definition as the corresponding class from the classical {\em Decorator} pattern -- the difference consists of the additional method {\tt\small getBase()} that returns the wrapped object. This method allows the fulfilment of the force no. 3 that allows decorations to be dynamically removed. In addition, there is an abstract class {\tt\small DecoComponent} that defines methods that correspond to the newly added interface-responsibilities. These methods have the role of dispatcher methods, meaning that they allow finding and calling the real implementations of the homonymous (methods with the same name) methods in the concrete decorators.

The definition of the {\tt\small Decorator} corresponds to a  {\em ForwardingDecorator} that has been proved to deal well with the fragile-base class problem (\cite{effective_java}), and also allows undecorated components to be used through {\tt\small DecoComponent} references (i.e. a {\tt\small ConcreteComponent} is just wrapped with a {\tt\small Decorator}).
\textbf{
	\begin{figure}
		\centering{
			\includegraphics[scale=0.45]{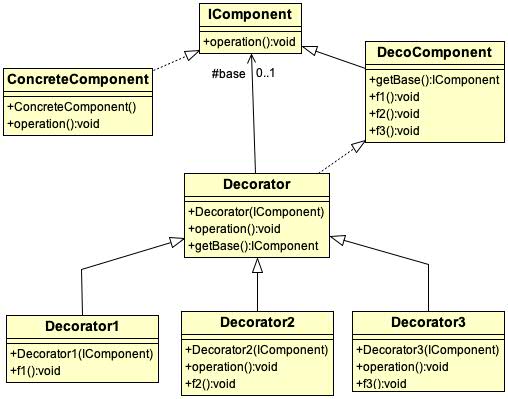}
		}
		\caption{The class diagram for the MixDecorator pattern.}
		\label{fig:mixdecorator}
	\end{figure}
}

In order to better explain  the pattern, we will give some implementation details in {\sf Java}. The {\tt\small Decorator} class is defined similarly to  the classical pattern, but it extends {\tt\small DecoComponent} (indirectly also {\tt\small IComponent}) and defines the method {\tt\small getBase()}  --Listing \ref{lst:mixdecorator}.

	\begin{figure}
	\centering{
\begin{lstlisting}[caption = {{\em MixDecorator} -- {\tt\small Decorator} class.},captionpos=b, label =lst:mixdecorator, frame=single, backgroundcolor=\color{light-gray}, basicstyle=\small\ttfamily, language=Java, numbers=left, numberstyle=\tiny             
     \color{black}]
public class Decorator implements DecoComponent {
	protected IComponent base;
	public Decorator(IComponent base) {
	   this.base = base;
	}
	public void operation() {
	   base.operation();
	}
	public IComponent getBase() {
	   return base;
	}	
} //~ end of class Decorator
\end{lstlisting}}
	\end{figure}

As it can be seen from Fig. \ref{fig:mixdecorator}, the concrete decorator classes {\tt\small Decorator1, Decorator2, Decorator3} are derived from {\tt\small Decorator} and implicitly from {\tt\small DecoComponent}. 
For a particular application/framework, after the new interface-responsibilities are inventoried, then the class {\tt\small DecoComponent} could be  defined.\\
 Very important is the fact that we must allow the posibility to extend the set of the methods defined inside {\tt\small DecoComponent}, and this  is analysed and discussed in section \ref{sec:mixdecorator.extensions}.

Since a method corresponding to a new interface-responsibility (as {\tt\small f1()}) could be defined into a decoration which is present somewhere in the chain of the decorations, we need a searching mechanism for calling this concrete method.
The role of the methods of {\tt\small DecoComponent} is to define this recursive searching mechanism. 

For Java implementation, we can define  {\tt\small DecoComponent} as an interface with default methods. A default method  is a virtual method that specifies a concrete implementation within an interface: if any class implementing
the interface will override the method, the more specific implementation
will be executed (\cite{DefaultMethods}).

The code snippet that corresponds to the {\tt\small DecoComponent} implementation in Java could be defined as it is shown in
the Listing \ref{lst:decorator_op}, where the implementation for the {\tt\small f1()} method is given.
The  code  hides a recursion that tries to call the method {\tt\small f1()}, and if this is not available for the top decoration, it  goes further to the previous decoration, by using {\tt\small getBase()}.
The recursion stops either when the concrete method is found or when it arrives to an undecorated component. 
The call of {\tt\small ((DecoComponent)base).f1()} could either lead to the invocation of the concrete implementation of {\tt\small f1()} -- iff the {\tt\small base} is exactly the decorator that defines {\tt\small f1()}, or  to another invocation  of the {\tt\small f1()} method defined into {\tt\small  DecoComponent}.

	\begin{figure}
	\centering{
\begin{lstlisting}[caption = {{\em MixDecorator} -- {\tt\small DecoComponent} in the context of {\em MixDecorator}},  captionpos=b, label =lst:decorator_op, frame=single, backgroundcolor=\color{light-gray}, basicstyle=\small\ttfamily, language=Java, numbers=left, numberstyle=\tiny             
     \color{black}]
public interface DecoComponent extends IComponent {
   default public void f1() throws UnsupportedFunctionalityException {
      IComponent base = getBase();
      // if base is a decorated object
      if (base instanceof DecoComponent) { 
         ((DecoComponent)base).f1();
      }
      // if base is not a decorated object
      else throw new UnsupportedFunctionalityException("f1");
   }
   ...
} //~ end of interface DecoComponent
\end{lstlisting}
}	\end{figure}

The following code snippet is an example that emphasizes the forces fulfillment; the execution throws no exception, and it can be noticed that, for example, {\tt\small f3()} and  {\tt\small f2()} could be called even if neither {\tt\small Decorator3} nor {\tt\small Decorator2} are the last added decoration.

\begin{figure}
\centering{
\begin{lstlisting}[ caption = {Testing different methods calls when the set of operations is extended},captionpos=b, label =lst:test_mixdecorator, frame=single, backgroundcolor=\color{light-gray}, basicstyle=\small\ttfamily, language=Java, numbers=left, numberstyle=\tiny             
\color{black}]
IComponent c = new ConcreteComponent();
DecoComponent dc = new Decorator(c);
dc.operation(); // Decorator just forward the call to ConcreteComponent
DecoComponent d = new Decorator1(new Decorator2(new Decorator3(c)));
d.operation();
d.f1(); d.f2(); d.f3();	
\end{lstlisting}
}\end{figure}

When the object {\tt\small d} invokes the method  {\tt\small f1()}, since {\tt\small Decorator1} overrides the method {\tt\small f1()}, the concrete implementation defined in {\tt\small Decorator1} is called. When method  {\tt\small f2()} is called, first its implementation from {\tt\small DecoComponent} is called, but then the call is sent forward to the base, which is the object obtained through ``{\tt\small new Decorator2(new Decorator3(c)}''; this call will invoke the definition of the method from {\tt\small Decorator2}. Similar mechanism is used when {\tt\small f3()} is called.

{\bf\em Remark:} \vspace{-0.1cm}
\begin{itemize}
	\item In this general solution the  {\tt\small DecoComponent} is defined as a separate class in order to specify the newly added responsibilities, and how are they treated. A much simpler solution is to combine it with the {\tt\small Decorator} class itself. 
\end{itemize}

\subsection{ Extensions with Other Responsibilities}
\label{sec:mixdecorator.extensions}
     
The design may imply a  dynamic development, and so new useful decorations could be  discovered in time, and these new decorations could define new interface-responsibilities, too.
To solve this problem it wouldn't be enough just to extend the {\tt\small DecoComponent} class to a class {\tt\small DecoComponent\_Extended} and define the new decorators by extending this class, because in this way we do not achieve fully compatibility between all the decorators.
 For example, if we considered {\tt\small Decorator4} such a decorator that defines the new method {\tt\small f4()} and we define in the class {\tt\small DecoComponent\_Extended} 
the method {\tt\small f4()} (similarly to {\tt\small f1()} from {\tt\small DecoComponent}), then the following code produces the correct execution for {\tt\small d41} but incorrect for {\tt\small d341}.

	\begin{figure}
	\centering{
\begin{lstlisting}[ caption = {{\em MixDecorator} -- Testing different methods calls when the set of operations is extended},captionpos=b, label =lst:test_mixdecorator_ext, frame=single, backgroundcolor=\color{light-gray}, basicstyle=\small\ttfamily, language=Java, numbers=left, numberstyle=\tiny             
     \color{black}]
IComponent c = new ConcreteComponent();
DecoComponent d1 = new Decorator1(c);
DecoComponent_Extended d41 = new Decorator4(d1);
d41.f1(); d41.f4(); //correct
DecoComponent d341 = new Decorator3(d41);   
d341.f4(); //incorrect -- error
\end{lstlisting}
}
\end{figure}


 For allowing extensions with new decorations that define interface-responsibilities, we need to be able to add new methods to the {\tt\small DecoComponent} interface/class and to provide  a basic implementation for them, too.
Many modern languages offer the possibility to define extension methods -- as {\sf C\#} (\cite{CSharp8}), or {\sf Kotlin} (\cite{kotlin}), or  some other similar mechanisms -- as default interface methods in {\sf Java} (\cite{DefaultMethods}). 

\subsection{Implementation Analysis and Possible Simplifications}
\label{sec:implementation}

The general solution of {\em MixDecorator} with the structure emphasized in Figure \ref{fig:mixdecorator}, could be implemented in any object-oriented language, but it is imperative to allow also extensibility and this comes with a special requirement:
\begin{itemize}
    \item  {\bf General requirement for {\em MixDecorator}:}
The language should provide the possibility of adding new methods to an interface/class, and  also to provide  a basic implementation for them.
\end{itemize}

\noindent By  using specific language constructs, the pattern could be simplified and also improved.\\
We will analyse this for several languages.

\subsubsection{{\sf Java} Implementation}
\label{sec:implementation.java}

As could been seen in the previous code snippets, the {\sf Java} solution is based on using interface with {\em default methods}.
The primary intent of introducing default methods in Java was to allow interfaces to be extended over time preserving backward compatibility. They are also associated to traits mechanisms as was proved by \cite{JavaTraits}.

	\begin{figure}
	\centering{
\begin{lstlisting}[caption = {{\em MixDecorator} -- {\sf Java} {\tt\small DecoComponent} class redefinition when new operations are added},  captionpos=b, label = {lst:java_redef}, frame=single, backgroundcolor=\color{light-gray}, basicstyle=\small\ttfamily, language=Java, numbers=left, numberstyle=\tiny             
\color{black}]
public interface DecoComponent extends IComponent {
   // ... all the other definitions
   default void f4() throws UnsupportedFunctionalityException {
      IComponent base = getBase();
      if (base instanceof DecoComponent)
         ((DecoComponent) base).f4();
      else throw new UnsupportedFunctionalityException("f4");
   }
} //~ end of interface DecoComponent
\end{lstlisting}
}\end{figure}

Based on this mechanism included in {\sf Java 8}, we may update the implementation of the interface {\tt\small  DecoComponent} by adding new methods that correspond to the newly added responsibilities (Listing \ref{lst:java_redef}). 
The initial decorators classes do not have to be recompiled, and no adaptation is needed. Still, we need to have access to the {\tt\small DecoCommponent} interface in order to replace its implementation, or at least to be able to specify the path where this is defined.
Considering this, the implementation of the method {\tt\small f4()} could  be added to {\tt\small  DecoComponent} as emphasized in Listing \ref{lst:java_redef}.
In this way, any decorators' combination is possible -- the newly added decorators could be "covered" by the previously defined ones, and vice-versa.

\subsubsection{{\sf C\#} Implementation}
\label{sec:implementation.csharp}

Using  {\em extension methods}  in {\sf C\#} we are able to add new methods to a class after the complete definition of the class (\cite{CS}).
They allow the extension of an existing type with new functionality, without having to sub-class or recompile the old type.

But, this mechanism allows only {\em static binding} and so the methods that could be added to a class cannot be declared virtual.
In fact, an extension method is a static method defined in a non-generic static class, but which can be invoked using an instance method syntax. 

The {\sf C\#} solution for extensibility requires the definition of a new static class that defines the extension methods for the {\tt\small DecoComponent} class (or directly to the {\tt\small Decorator} class).  The extension methods define the recursive search mechanisms for the new methods. What is different in the  {\sf C\#} solution is that being based on static methods, the base case should be also treated inside the extension method. The base case is represented by the situation when the invoked responsibility is defined by a method of the last added decorator.


For each new interface-responsibility or for a set of interface-responsibilities, a new static class could be defined -- this class defines extension methods that specify the recursive search mechanisms.

More concretely, we may add a static class {\tt\small Decorator\_Extension} where the method {\tt\small f4()} is defined as extension method. The class {\tt\small Decorator\_Extension}  provides extension for {\tt\small DecoComponent}:

	\begin{figure}
	\centering{

\begin{lstlisting}[caption = {{\em MixDecorator} -- {\sf C\#} definition of the extension methods for the {\tt\small Decorator}  class}, 
 captionpos=b, 
label = {lst:Csharp_mixdecorator},
frame=single,  backgroundcolor=\color{light-gray}, basicstyle=\small\ttfamily, language=Java, numbers=left, numberstyle=\tiny             
\color{black}]
public static class Decorator_Extensions {
	public static void f4(this DecoComponent cdb) {
		Decorator4 cdb4 = cdb as Decorator4;
		if (cdb4 != null) cdb4.f4();  
		else {
			DecoComponent cdb_base = cdb.getBase() as DecoComponent;
			if (cdb_base != null) cdb_base.f4();       
			else throw new UnsupportedFunctionalityException("f4");   
		}
	}
} //~ end of class Decorator_Extension
\end{lstlisting}
}\end{figure}

In {\sf Java}, the base case of the recursive search  is implicitly done based on polymorphic call, but when {\sf C\#} extension methods are used, this case should be explicitly defined.

In {\sf C\#} the simplification could be done by defining extension methods directly to the {\tt\small Decorator} class, and by excluding  {\tt\small DecoComponent}.  

\subsubsection{{\sf C++} Implementation}
\label{sec:implementation.c++}

In a language like C++, we don't have extension methods, but  we still have to respect the constraint of allowing  the definitions of new interface-responsibilities.  
To overcome this, we may use template classes with policies in order to postpone the specification of the parent class for decorators. C++ Policies could be considered a very  interesting and useful meta-programming mechanism that allows behavior infusion in a class through templates and inheritance as was analysed by \cite{Alexandrescu2001ModernCD,metaprogramming04}. On the other hand, \cite{C++Mixins} show that in C++, mixins could be defined using templates, and we may consider that  C++ mixins could be implemented using policies. 



 The solution is to force the {\small\tt Decorator} class to extend the most recently defined {\tt\small DecoComponent} class. 
So, the {\small\tt Decorator } class is defined as a template class, and  the template parameter will also be used as a parent class for the {\small\tt Decorator } class -- Listing \ref{lst:c++_mixdecorator}. 
This is necessary because, in this way, we may postpone the specification of the base class, and allow this base to be either {\tt\small DecoComponent} or another class that extends {\tt\small DecoComponent} (e.g. {\tt\small DecoComponent\_Extended}). 

	\begin{figure}
	\centering{
\begin{lstlisting}[caption = {{\em MixDecorator} -- {\tt\small Decorator} class definition in {\sf C++}},captionpos=b, label={lst:c++_mixdecorator}, frame=single, backgroundcolor=\color{light-gray},language=C++, basicstyle=\footnotesize\ttfamily, numbers=left, numberstyle=\tiny             
     \color{black}]
template <typename T=DecoComponent>
class Decorator: public T {
protected:
    IComponent* base;
public:
    Decorator(IComponent* base) {
        this->base = base;
    }
    void operation() {
        base->operation();
    }
    IComponent* getBase() {
        return base;
    }
};
\end{lstlisting}
}\end{figure}

The concrete decorators are also  defined as template classes, since they are derived from the {\tt\small Decorator} class.
{\tt\small DecoComponent} would still be used as a parent class for the {\tt\small Decorator} class, but this relation will be defined through the template parameter.
The {\tt\small DecoComponent} class defines the corresponding search methods for the newly defined  methods  in the concrete decorators (Listing \ref{lst:c++_decocomponent}).  

	\begin{figure}
	\centering{
\begin{lstlisting}[caption = {{\em MixDecorator} -- {\sf C++} implementation of {\tt\small DecoComponent} and its method {\tt\small f1} },captionpos=b, frame=single, backgroundcolor=\color{light-gray}, basicstyle=\footnotesize\ttfamily, numbers=left,  label = {lst:c++_decocomponent},language=C++,
numberstyle=\tiny \color{black}]
class DecoComponent: public IComponent {
public:
   virtual IComponent* getBase() = 0;
   virtual void f1();
   virtual void f2();
   virtual void f3();
};
void DecoComponent::f1() {
   IComponent* base = getBase();
   DecoComponent* decor = static_cast<DecoComponent *>(base);
   if (decor) { // if is decorated
      decor->f1();
   }
   else
      throw new UnsupportedFunctionalityException("f1");
}
// similar definitions for f2 and f3
\end{lstlisting}
}\end{figure}

The class {\tt\small DecoComponent} could be extended (e.g.  {\tt\small DecoComponent\_Extended}) with classes that define new methods that correspond to the new responsibilities added into additional decorators (Listing \ref{lst:c++_decocomponent_extended}). 
	\begin{figure}
	\centering{
\begin{lstlisting}[caption = {{\em MixDecorator} -- {\sf C++} implementation of {\tt\small DecoComponent\_Extended} and its method {\tt\small f4} },
captionpos=b, frame=single, backgroundcolor=\color{light-gray}, basicstyle=\footnotesize\ttfamily, numbers=left,  label = {lst:c++_decocomponent_extended},language=C++,
numberstyle=\tiny \color{black}]
class DecoComponent_Extended: public DecoComponent {
public:
   virtual void f4();
};

void DecoComponent_Extended::f4() {
   IComponent* base = getBase();
   DecoComponent_Extended* decor = static_cast<DecoComponent_Extended *>(base);
   if (decor) { // if is decorated
      decor->f4();
   }
   else
      throw new UnsupportedFunctionalityException("f4");
}
\end{lstlisting}
}\end{figure}

When the decorators are used, we need to specify their parent class that could be either {\tt\small DecoComponent} or {\tt\small DecoComponent\_Extended}. In this way, all the new responsibilities are correctly found. 

The  example presented in Listing \ref{lst:c++_test_mixdecorator} shows first  a decoration with {\tt\small Decorator1} (for which we have the correspondent class {\tt\small DecoComponent}), and then we have an object decorated with {\tt\small Decorator1}, which is added over {\tt\small Decorator4}; in this case we have to specify  {\tt\small DecoComponent\_Extended} as a superclass (template parameter). Since {\tt\small DecoComponent\_Extended}  class extends {\tt\small DecoComponent}
 the function {\tt\small f4()} could be called even after {\tt\small Decorator1} was added, and {\tt\small f1()} is also accessible.

	\begin{figure}
	\centering{
\begin{lstlisting}[caption = {{\em MixDecorator} -- Testing different methods' calls in {\sf C++} implementation},captionpos=b,  frame=single, label = lst:c++_test_mixdecorator, backgroundcolor=\color{light-gray},language=C++, basicstyle=\footnotesize\ttfamily, numbers=left, numberstyle=\tiny     \color{black}]
IComponent* c = new ConcreteComponent();
Decorator<DecoComponent>* dc = new Decorator1<DecoComponent>(c);
dc->f1();
Decorator<DecoComponent_Extended>* dc14 = 
   new Decorator1<DecoComponent_Extended>(
    new Decorator4<DecoComponent_Extended>(c));
dc14->f4();
dc14->f1();
dc14->operation();
\end{lstlisting}
}\end{figure}

\noindent {\bf\em Remarks:} 

\begin{itemize}
\item This solution used for C++, emphasized also an interesting mechanism that allows creating a new class from different base classes, and so adjust the type that we need using the inheritance mechanism. It could be seen also as a mixin mechanism (\cite{mixin}). 
\item Since in {\sf C++} the template instantiation leads to the creation of a new class it is a statical approach, but only for the supertype specification.
\end{itemize}

\subsubsection{{\sf Python} Implementation}
\label{sec:implementation.python}

In {\sf Python}  a classical decorator pattern approach could be used, together with a searching mechanism method, which could be easily specified using {\tt\small \_\_getattr\_\_()} method -- as it is emphasized by the  code presented in Listing \ref{lst:python_icomponent}.

	\begin{figure}
	\centering{
\begin{lstlisting}[caption = {{\em MixDecorator} -- {\sf Python} implementation  of the components and decorators}, captionpos=b, frame=single, backgroundcolor=\color{light-gray}, basicstyle=\footnotesize\ttfamily,label = {lst:python_icomponent}, numbers=left, language=python, numberstyle=\tiny             
\color{black}]
class IComponent:
   __metaclass__ = ABCMeta
   @abstractmethod
   def f(self): pass
   
class ConcretComponent(IComponent):
   def f(self):
      print("original f")
      
class Decorator(IComponent):
   def __init__(self, decoratee):
      self._decoratee = decoratee
   def f(self):
      print("decorated f")
      self._decoratee.f()
   def __getattr__(self, name):
      return getattr(self._decoratee, name)
      
class Decorator1(Decorator):
   def __init__(self, decoratee):
      self._decoratee = decoratee
   def f1(self):
      print("original f1")
      
# the definition of Decorator1, Decorator2 and Decorator3 are similar
\end{lstlisting}
}\end{figure}

\noindent If one needs a new decorator {\small\tt Decorator4} that defines a new method ({\small\tt f4}), a new class could be 
directly derived from the {\small\tt Decorator} class. 
Using this approach all the methods could be used in any order, as emphasised in the Listing \ref{lst:python_decorator_ext}:
	\begin{figure}
	\centering{
\begin{lstlisting}[caption = {{\em MixDecorator} -- {\sf Python}  decorators that define new methods, and their testing}, captionpos=b,
 label = {lst:python_decorator_ext},frame=single, backgroundcolor=\color{light-gray}, basicstyle=\footnotesize\ttfamily, numbers=left,language=python, numberstyle=\tiny             
\color{black}]
class Decorator4(Decorator):
   def __init__(self, decoratee):
      self._decoratee = decoratee
   def f4(self):
      print("original f4")
      
c = ConcretComponent()
d = Decorator1(Decorator4(Decorator3(c)))
d.f()
d.f1()
d.f4()
\end{lstlisting}
}\end{figure}

The {\tt\small  \_\_getattr\_\_()} method is implicitly used to lookup for the class attributes, being used to define  the recursive search.

\subsubsection{Comparison of the implementation solutions}
\label{sec-mixdecorator.comparison}

We may notice that the solutions are simpler if the language provides mechanisms that allow some kind of type extensions, such as mixins or traits.

Probably the simplest solution is the one that {\sf Python} offers, but this is due to the fact that {\sf Python} is a dynamical typed language.

   At a glance, {\sf C\#} provides also a very simple and nice solution, but still there are situations when it could become very complex because it is based on a {\em static} mechanism.     
    The solution of C\# implementation based on extension methods could avoid the definition of the class {\tt\small DecoComponent} since all the extensions could be done on the same interface {\tt\small IComponent} (but semantically this is maybe not the best solution). 
		A definition of a new decorator that defines a new interface-responsibility, even in a further step  of development, does not imply more operations  for a decorator that has been defined at the first step of the development.
     
     The C\# extensions methods are static methods that are called as they are instance methods. In principle, this static character does not affect the solution, because, in this case, these static methods provide just a search mechanism for the instance methods with the same name. This search mechanism  does not have to be changed dynamically.
      Still, if there are more decorations that define new responsibilities but with the same name (for example there is also a {\tt\small Decorator4\_prime} that extends {\tt\small Decorator4} and overrides the method {\tt\small f4()}), the {\sf C\#} solution should verify in chain all the possibilities starting from the most specialized class (i.e. calling {\tt\small f4} from {\tt\small Decorator4\_prime} or from {\tt\small Decorator4}). The  code snippet in Listing \ref{lst:c++_decocomponent_ext_problem} emphasizes such a situation.
     
     	\begin{figure}
     	\centering{
                 \begin{lstlisting}[caption = {{\em MixDecorator} -- {\sf C\#} extension methods that correspond to an operation defined in two decorators}, captionpos=b, frame=single,  label = {lst:c++_decocomponent_ext_problem}, backgroundcolor=\color{light-gray}, basicstyle=\footnotesize\ttfamily, language=Java, numbers=left, numberstyle=\tiny             
     \color{black}]
public static class DecoComponent_Extended4 {
	public static void f4(this DecoComponent cdb) {
		Decorator4 cdb4 = cdb as Decorator4_prime;
		if (cdb4 != null) cdb4.f4();  
		else {
			cdb4 = cdb as Decorator4;
			if (cdb4 != null) cdb4.f4();  
			else {
				DecoComponent cdb_base = cdb.getBase() as DecoComponent;
				if (cdb_base != null) cdb_base.f4();       
				else throw new UnsupportedFunctionalityException("f4");   
			}
		}
	}
} //~ end of class DecoComponent_Extended4
\end{lstlisting} 
}
\end{figure}     

 Since the Java solution is based on polymorphic calls, this problem doesn't appear in the Java implementation.
     
 Based on these, we may conclude that apparently the implementation of {\em MixDecorator} pattern is simpler and easier with {\sf C\#} extension methods. But based on the previous analysis, it would be recommended to use default methods in C\#, too; they were just added in C\# 8.0 (\cite{CSharp8}). The implementation based on default methods is more object-oriented rigorous and more efficient since some actions are done implicitly based on polymorphic call of the invoked methods. 

The C++  also provides a good solution for {\em MixDecorator}  implementation; this is based on templates, which represents static mechanisms, too. Even though, the C++ solution does not have the C\# problems since the methods ({\tt\small f1(), f2(), ...}) are defined as virtual functions.

\subsection{ {\em MixDecorator} Consequences}
\label{sec:mixdecorator.consequences}

The solution offered by the {\em MixDecorator} pattern preservers the general advantages of {\em Decorator}, but  presents several new advantages:

\begin{itemize}
	
	
	\item The linear combination of the decorations is hidden. The final object could be seen as an object with a set of additional responsibilities. 
	
	\item It is easy to add any combination of capabilities. The same capability can even be added twice. 
	
	\item Added behaviour could be used in any combination, without any additional operation, such as withdrawing decorations. 
	
	\item Clients have to refer to {\tt\small DecoComponent} interface through which concrete or decorated  components can be used.

\item The definition of {\tt\small DecoComponent}  implies a specification of all new interface-responsibilities that are defined through decorations; even if initially just a small set is defined, it could be extended with an adaptation that is language dependent.  

\end{itemize}


\section{{\em D2Decorator} -- A Variant Based on Double-Dispatch}
\label{sec:double-dispatch}

The {\em MixDecorator} solution presented in  section \ref{sec:mixdecorator} collects together the methods corresponding to different interface-responsibilities and provides for each a recursive search mechanism inside the interface/class {\tt\small DecoComponent}. When new decorators are defined, {\tt\small DecoComponent} should be extended with new methods. 
Another way of solving the messages calls is to use a double-dispatching mechanism.

\subsection{{\em D2Decorator} Solution}
\textbf{
	\begin{figure}
		\centering{
			\includegraphics[scale=0.45]{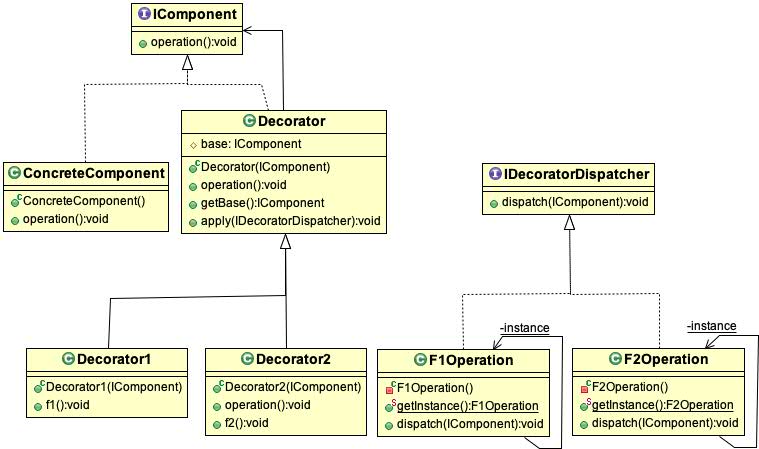}
		}
		\caption{The class diagram for the D2Decorator pattern.}
		\label{fig:d2decorator}
	\end{figure}
}
Instead of defining an all-operations interface, we may define for each newly added decorator, which defines a new interface-responsibility, a specialized dispatcher that could be used in the search  mechanism. The dispatchers are objects that have the responsibility of executing their corresponding methods.
The recursive search algorithm is extracted in only one  method that is parameterised with an argument of dispatcher type.

The structure of {\em D2Decorator} pattern is shown in  Fig. \ref{fig:d2decorator}.
The solution includes the structure of the classical {\em Decorator}, but needs to use also  special dispatcher classes. The interface {\tt\small IDecoratorDispatcher} defines one method, {\tt\small dispatch}, which will be implemented by the special classes -- the dispatchers -- that have the role to execute the newly added interface-responsibilities. It can be seen as a functional interface.

	\begin{figure}
	\centering{
\begin{lstlisting}[caption = { {\em D2Decoreator} -- the method {\tt\small  apply()} from {\tt\small Decorator class}},captionpos=b, frame=single, label=lst_d2decorator_apply, backgroundcolor=\color{light-gray}, basicstyle=\footnotesize\ttfamily, language=Java, numbers=left, numberstyle=\tiny             
     \color{black}]
public class Decorator extends IComponent {
   protected IComponent base;    
   // ...
   public void apply(IDecoratorDispatcher d) throws
      UnsupportedFunctionalityException {
      try {   
         d.dispatch(this); // base case   
      }
      catch (UnsupportedFunctionalityException  e) {
         if (base instanceof Decorator)
            ((Decorator) base).apply(d); // the recursive call 
		 else throw e;
	  }
   }
} //~ end 
\end{lstlisting}}
\end{figure}

The search mechanism is again recursive, and relies on type casting and extracting the wrapped object -- {\tt\small base}. The  link with the method that should be called is done through the dispatcher associated to that method.
The class {\tt\small Decorator} provides a general execution operation {\tt\small apply} that receives an argument of type {\tt\small IDecoratorDispatcher}. The method asks first the received dispatcher to try to execute the desired operation through the {\tt\small dispatch} method, and if it does not succeed then,  if the {\tt\small base} is also a decorator, the {\tt\small apply} method is recursively invoked  for the {\tt\small base} -- Listing \ref{lst_d2decorator_apply}.
The {\tt\small dispatch } method invoked by the dispatcher argument will call the associated operation defined in the corresponding {\tt\small Decorator}.

For {\tt \small Decorator1} that defines the method {\tt\small f1()}  a dispatcher {\tt\small F1Operation} should be defined as the code in Listing \ref{fst:d2decorator_F1Operation} emphasizes:

	\begin{figure}
	\centering{
\begin{lstlisting}[caption = { {\em D2Decoreator} -- The method {\tt\small dispatch} inside a concrete implementation of {\tt\small IDecoratorDispatcher} interface: {\tt\small F1Operation} }, captionpos=b, frame=single,
label=fst:d2decorator_F1Operation,
backgroundcolor=\color{light-gray}, basicstyle=\footnotesize\ttfamily, language=Java, numbers=left, numberstyle=\tiny             
     \color{black}]
public class F1Operation implements IDecoratorDispatcher {
	public void dispatch(IComponent c) throws UnsupportedFunctionalityException {
		if (c instanceof Decorator1)
			((Decorator1)c).f1();
		else throw new UnsupportedFunctionalityException("f1");
	}
} //~ end of class F1Operation
\end{lstlisting}
}\end{figure}

{\bf\em Remark:} Since, for an operation it is enough to have only one instance of the corresponding dispatcher, the {\em Singleton} pattern could be used for the dispatchers.

\vspace{0.3cm}
\noindent All the operations could be called, regardless the order in which the decorators were added. 
The following code snippet emphasizes the call of operation {\tt\small f1()} followed by the call of operation {\tt\small f2()}:

	\begin{figure}
	\centering{
\begin{lstlisting}[caption = {{\em D2Decoreator} --  Testing different methods calls. }, captionpos=b, frame=single,
label=lst:d2decorator_testing,
backgroundcolor=\color{light-gray}, basicstyle=\footnotesize\ttfamily, language=Java, numbers=left, numberstyle=\tiny             
\color{black}]
Decorator d12 = new Decorator1(new Decorator2(new ConcreteComponent()));

d12.operation();
d12.apply(F1Operation.getInstance());
d12.apply(F2Operation.getInstance());
\end{lstlisting}
}\end{figure}

It can be noticed that the calls of the methods that do not belong to the {\tt\small IComponent} interface, are done through the method {\tt\small apply}, and the differentiation is specified by the type of parameter received. Each dispatcher type is connected to one method that represents an interface-responsibility.

\subsection{{\em D2Decorator} Implementation}

The given examples do not show the case when the new methods also have arguments. If the new methods don't have arguments then any object-oriented language could be used for the implementation.

In order to allow arguments for the newly added responsibilities,
we must define the function {\tt\small apply} having beside the dispatcher argument, a variable list of arguments; this list has to be then sent to the {\tt\small dispatch} method, which in turn will use it for calling  the actual method. 

From this we deduce that the most important requirement for a general {\em D2Decorator} implementation is related to the need of having  methods with  variable lists of arguments.

\begin{itemize}
\item {\bf General requirement for {\em D2Decorator}:}
The requirement of a language to support the {\em D2Decorator} implementation is  the need to define variable lists of arguments and the possibility to transfer this list to another function call.
\end{itemize}

\subsection{{\em D2Decorator} Consequences}
As {\em MixDecorator}, the {\em D2Decorator} pattern preserves the benefits of the {\em Decorator} pattern, but additionally comes with some advantages and disadvantages.

\noindent Advantages:
\begin{itemize}
\item We may mix together new and old decorators that define new interface-responsibilities, in any order, and all methods that define interface-responsibilities are  accessible. 
\item In contrast with {\em MixDecorator}, the recursion for finding the invoked method is defined only once in the method {\tt\small apply()}. This solution factorizes the invocation process by separating the recursive search definition by the actual function call.
\end{itemize}

\noindent Disadvantages:
\begin{itemize}
\item An important disadvantage is given by the fact that in order to call the new interface-responsibilities, we have to use the method {\tt\small apply()} instead of a simple invocation of the corresponding method by its name.

\item  For each newly added interface-responsibility, a new dispatcher class should be defined; each decorator that defines new interface-responsibilities has to come together with corresponding dispatchers for each such an interface-responsibility. This could lead to a lot of fine-grain classes.

\item The {\em D2Decorator} works well and it is also simple iff the new methods don't have arguments.  When the method that should be called has arguments, these should be taken from the argument list of the {\tt\small apply} method. These arguments are taken from the {\tt\small apply} method and sent through the {\tt\small dispatch()} method to the actual method.
Depending on the implementation language this could become a difficult task.

\end{itemize}

\section{Static versus Dynamic solution - {\em HybridDecorator}}
\label{sec:hybrid}

For the {\em MixDecorator} C++ implementation,
we have shown how the template mixins (policies)
could be used as a mean for enlarging the interfaces.
Template policies could also be used
for defining another alternative, a  hybrid solutions of the {\em Decorator}
pattern. 

Templates are metaprogramming mechanisms that allow macro-definitions from which new classes are created at the compilation phase; so they are static mechanisms. The decorators are defined as mechanisms that allow dynamic adaptation of the objects -- they allow adding responsibilities dynamically and transparently, and withdrawing these responsibilities when  they are not longer necessary. 

We propose a variant for which the combinations of the added responsibilities are defined statically, but they could be added and removed dynamically to/from an object.


The presented solution -- {\em HybridDecorator} -- is based on C++ policies that are
implicitly based on inheritance; but the idea that inheritance
should be avoided was in the context of using
inheritance for creating all possible combinations of decorations
-- and this is not the case for this solution. The
recursion implicitly involved by the templates policies
fits very well to the recursion implied by the Decorator
definition. (This variant was first proposed by \cite{Nic20}.)

\subsection{{\em HybridDecorator} solution}

The hybrid solution  is both static and dynamic, and also uses both inheritance and composition.  This  preserves the basic object wrapping, but the decorations will be defined as a single class obtained  through a chain of inheritance derivations. 
In order to assure this, we need to define a class {\tt\small Decorator} that provides the support for composition, but at the same time, intermediates the decorations inheritance.

\begin{figure}
	\centering{
		\includegraphics[scale=0.65]{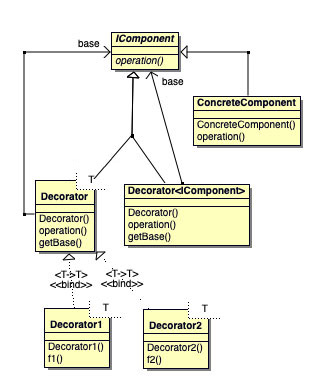}
	}
	\caption{The class diagram of the {HybridDecorator} pattern.}
	\label{fig:HybridDecoratorDiagram}
\end{figure}

\noindent The class is defined for the following two cases: 

\begin{itemize}
	\item the basic case with 
	the template parameter equal to {\tt\small IComponent}, 
	\item  the general case with a general template parameter({\tt\small T}).
\end{itemize}

	\begin{figure}
	\centering{
\begin{lstlisting}[caption = {{\em HybridDecorator} -- The definition of the class {\tt\small Decorator}    for the general template parameter },captionpos=b, label=lst:hybridDecorator, frame=single, backgroundcolor=\color{light-gray}, basicstyle=\footnotesize\ttfamily, numbers=left, numberstyle=\tiny             
\color{black}, language=C++]
template <typename T>
class Decorator: public T {
protected:
   IComponent* base;
public:
   Decorator(IComponent* r): T(r) {
      this->base = r;
   }
   void operation() {
      T::operation();
   }
   IComponent* getBase() {
      return T::getBase();
   }
};
\end{lstlisting}
}\end{figure}

	\begin{figure}
	\centering{
\begin{lstlisting}[caption = {{\em HybridDecorator} -- The definition of the class {\tt\small Decorator}   for the implicit template parameter {\tt\small IComponent}},captionpos=b, label=lst:HybridDecoratorImplicit, 
frame=single, backgroundcolor=\color{light-gray}, basicstyle=\footnotesize\ttfamily, numbers=left, numberstyle=\tiny             
\color{black}, language=C++]
template <>
class Decorator<IComponent>: public IComponent {
protected:
   IComponent* base;
public:
   Decorator(IComponent* r) {
      this->base = r;
   }
   void operation(){
      base->operation();
   }
   IComponent* getBase(){
      return base;
   }
};
\end{lstlisting}
}\end{figure}

These two cases are different, and this difference is mainly related to the call of the overridden method {\tt\small operation()}:
\begin{itemize}
	\item for the general case the call is sent up to the superclass -- {\tt\small T} -- Listing \ref{lst:hybridDecorator}.
	\item for the basic case the call is sent to the wrapped object ({\tt\small base}) -- Listing \ref{lst:HybridDecoratorImplicit};
\end{itemize}

The associated UML diagram for this hybrid implementation is shown in Fig. \ref{fig:HybridDecoratorDiagram}. The solution preserves the classical solution of using both inheritance and composition: {\tt\small Decorator} class ``has-a'' {\tt\small IComponent} (the attribute {\tt\small base}), but it is also derived from {\tt\small IComponent}. Still, as the diagram emphasizes, there are two definitions of the class {\tt\small Decorator}: one for the implicit type parameter {\tt\small IComponent}, and the other for the general case, when the template parameter could be any decorator specialization. The class {\tt\small Decorator} is derived from the template parameter, too. This is not visible in the UML diagram but it is important for the recursive definition of the decorators.
Also, the method {\tt\small getBase()} is necessary for this variant.

The concrete decorators -- as {\tt\small Decorator1} and  {\tt\small Decorator2} -- are derived from the class {\tt\small Decorator}, and so these are also template classes. For them it is not necessary to give special definition for the implicit type case ({\tt\small IComponent}) since this is solved in the {\tt\small Decorator} class. The definition of the  {\tt\small Decorator1} class is given  in  Listing \ref{code:Decorator1}.

	\begin{figure}
	\centering{
\begin{lstlisting}[caption = {{\em HybridDecorator} -- The definition of {\tt\small Decorator1} class. },captionpos=b, label=code:Decorator1, frame=single, backgroundcolor=\color{light-gray}, basicstyle=\footnotesize\ttfamily, numbers=left, numberstyle=\tiny             
\color{black}]
template <typename T=IComponent>
class Decorator1: public Decorator<T> {
public:
	Decorator1(IComponent* r): Decorator<T>(r) {
		Decorator<T>::base = r;
	}
	void f1() {   
		std::cout<<"call f1 in Decorator1"<<std::endl;
	}
};
\end{lstlisting}
}\end{figure}

An usage example based on this solution is given in the Listing \ref{code:exampleHybrid}. It can be noticed that the usage looks very similar to the classical {\em Decorator} implementation, but the composition is replaced with the template parameter specification.

	\begin{figure}
	\centering{
\begin{lstlisting}[caption = {{\em HybridDecorator} -- Usage example.},captionpos=b, label=code:exampleHybrid, frame=single, backgroundcolor=\color{light-gray}, basicstyle=\footnotesize\ttfamily, numbers=left, numberstyle=\tiny             
\color{black}]
// ConcreteComponent  instantiation
ConcreteComponent* c = new ConcreteComponent();

// call of the ConcreteComponent operation
c->operation();

// decorated component instantiation
Decorator3<Decorator1<Decorator2<>>> *deco =
//IComponent is the implicit value for the template parameter
   new Decorator3<Decorator1<Decorator2<IComponent>>>(c);

// calling the 'operation' through decorated component
deco->operation();

// calling the new interface-responsibilities
deco->f1();
deco->f2();
deco->f3();

// unwrap the object
IComponent* cc = deco->getBase();
// call again the ConcreteComponent operation
cc->operation();
\end{lstlisting}}\end{figure}

Through the  definition {\tt\small Decorator3<Decorator1<Decorator2<IComponent>>>}, a new class is created, and this class is obtained by deriving {\tt\small Decorator2} from {\tt\small IComponent}, {\tt\small Decorator1} from {\tt\small Decorator2}, and  {\tt\small Decorator3} from {\tt\small Decorator1}. This class has an attribute -- {\tt\small base} -- of type {\tt\small IComponent}. This specific inheritance is created only for this particular example; other variations are created when they are needed.

\subsection{ {\em HybridDecorator} Consequence} 

{\em HybridDecorator} presents several advantages and disadvantages:
\label{sec:hybriddecorator.consequences}

\noindent  Advantages:
\begin{itemize}
	\item Only the classes for the needed combinations are created.
	\item All new defined methods in different decorators are accessible.
	\item The combination of the functionalities could be changed  -- the basic object is retrieved (through the method {\tt\small getBase()}), and then it could be passed to another combination of functionalities.
	\item  For all the specializations of {\tt\small IComponent} only one particular class that corresponds to a particular functionality combination could be used;
	\item In order to add a new decoration (or combination), a new object wrapping (similar to the classical {\em Decorator}) could be done -- the wrapping will specify a new class with the desired decoration.
	\item static class creation assures efficiency,
	\item in comparison to the solutions of the {\em MixDecorator} or {\em D2Decorator},  {\em HybridDecorator} implicitly assures full accessibility of all new interface-responsibilities, without being necessary to define another special class/es or methods  to intermediate this. 
\end{itemize}

\noindent Disadvantages:
\begin{itemize}
	\item The decorations could not be individually  retrieved; 
	
	\item There is no supertype for all possible decorators' combinations to be used as a general reference type.
	\item It could be used by adding successively the decorators without using template policies, and in this case, the constraints related to accessibility are the same as for the  classical {\em Decorator}.
\end{itemize}

\vspace{0.2cm}

The hybrid aspect of this solution is very important: we have an object wrapped into a decoration object (composition), but this decoration object is defined as a combination of fine grain decorators that are built through inheritance defined using the C++ template mechanism. If only inheritance would be used (as it is emphasized by the code in the Listing \ref{lst:inheritancetdecorator}), then we wouldn't satisfy the {\em Decorator} forces that impose decorations to be dynamically added and withdrawn.

	\begin{figure}
	\centering{
\begin{lstlisting}[caption = {A variant based only on inheritance -- usage example },captionpos=b, label=lst:inheritancetdecorator, frame=single, backgroundcolor=\color{light-gray}, basicstyle=\footnotesize\ttfamily, numbers=left, numberstyle=\tiny             
 \color{black}]
Decorator1< Decorator2< Decorator3<ConcreteComponent>>> *d123c = 
  new   Decorator1<Decorator2<	Decorator3<ConcreteComponent>>>();
  			
d123c->f1(); d123c->f2(); d123c->f3(); d123c->operation();
 \end{lstlisting}}\end{figure}

{\em\bf Remark} Even if this variant is specific only to the {\sf C++} language, this  brings an interesting insight about the possibility to define the {\em Decorator} pattern using metaprograming mechanisms.

\section{Analysis and Conclusions}
\label{sec:conclusions}

The paper presents a new modern view over a classical and very used pattern -- {\em Decorator}.  An important restriction of the {\em Decorator} classical solution is the limitation to the component interface. This could be overcome by using pattern adaptations, which are being based on new developments of the object-oriented languages (default methods in interfaces, extension methods, template mixins, variable argument list), or on metaprogramming.

The presented variants: {\em MixDecorator}, {\em D2Decorator}, and {\em HybridDecorator} represent variants of the {\em Decorator} pattern since they allow modularized functionality specialization, but also, in addition, they facilitate the addition of new interface-responsibilities. In this way, the set of messages that could be sent to an object is enlarged. We may consider that using them, we {\em dynamically modify the type of an object}. Different combinations of these messages could be used, and all the responsibilities are directly accessible.


For the {\em MixDecorator}, the implementation constraint for extensibility is related to the fact that we have to be able to add a set of operations to an interface (or a class), and also to provide a basic implementation for the corresponding methods. This could be achieved by using language specific mechanisms, such as that provided by the Java extended interfaces, or the {\sf C\#} and {\sf Kotlin} extension methods. In {\sf C++} we may use an implementation variant based on mixin templates.

If we deeply analyze the {\em MixDecorator} solution, we may notice that the structure of the search algorithm for the concrete implementation of an  interface-responsibility is the same for all of them. So, it would be useful if we can generate the dispatcher methods based on some metaprogramming mechanisms, since they are particularized only by  the decorator type that define the new method, and the method's name and arguments.

The variant based on double-dispatch  -- {\em D2Decorator} --  factorizes the definitions of the dispatcher methods in different classes (the dispatchers), but the recursion has to be defined only in the {\tt\small Decorator} class. This variant could be applied in any object-oriented language, that allows variable list arguments,
offering the advantage of mixing together new and old decorators in any order, without adaptation or  modification. 
Still, its main disadvantage is that the methods should be called, not directly, but through a general method, {\tt\small apply(IDecoratorDispatcher d)} (and this also implies that for each interface-responsibility a new dispatcher class should be defined).

In a pattern oriented analysis, comparing with {\em Factory Method} and {\em Abstract Factory},  we may consider that we have defined and used {\em Dispatcher Method} and {\em Abstract Dispatcher} patterns.

{\em HybridDecorator} is specific only to C++, but it emphasizes an interesting 
solution of the problem for which the {\em Decorator} was initially proposed.
The inheritance is used for creating the combination of decorators, but this is done through the meta-programming mechanism brought by the C++ templates; in this way the desired decoration combination is created only when it is needed.

{\bf Mixins and traits oriented analysis}.
Traits and mixins are both related -- they allow the injection of some code into a class. Both constructs exploit composition instead of inheritance as a mechanism for software reuse and they are alternatives to multiple inherintance.
Since the applicability of the discussed {\em Decorator} variants is related to the possibility of adding new functionalities, a comparison with them it's worth to  be done.

{\em Mixin} programming is a style of software development where units of functionality are created in a class and then mixed in with other classes (\cite{mixin}). 
A mixin class could be considered as a parent class that is inherited from -- but this is not done in order to obtain a specialization. Typically, the mixin will export services to a child class, but no semantics will be implied about the child ``being a kind of'' the parent. 
The main differences between 
the {\em Decorator} pattern and {\em mixins} are based on the fact that with decorators we want to add functionality to objects, not to create new classes that contain a combination of methods from other classes.
With decorators we may extend functionality (change behaviour and add new responsibilities) of an object, and this new functionality could be added and removed dynamically. 
This could bring advantages for implementations in languages where there are no specific {\em mixins} mechanisms.
On the other hand, because for the {\em MixDecorator} we have to be able to add methods to the {\tt\small DecoComponent} (or directly to the {\tt\small Decorator} class), we may consider that the {\em MixDecorator} implementation could be based on mixins, if they are available in the implementation language (as it is the case of the C++ implementation).


%


Traits also allow the programmer to create components that are designed for reuse, rather than for instantiation (\cite{Schrli2002TraitsCU}).
Being stateless, they are more lightweight entities that serve as the primitive units of code reuse.
Java 8 extended interfaces that allow default method definitions may be considered a kind of trait mechanism -- 
Java Traits, as was analysed by \cite{JavaTraits}. We have used them in order to define the {\tt\small DecoCompoment} interface for the {\sf Java} {\em MixDecorator} implementation.
In {\em Scala} it is  possible to add  a trait to an object instance when the object is created, and not only to an entire class; but this is possible because in {\sf Scala} we work with singleton objects (\cite{ScalaTraits}). Adding functionality to an object is what we do using decorators.


\vspace{0.3cm}

As the examples with the {\tt\small Input/OutputStream} and {\tt\small ReaderDecorator} show, 
the applicability of these enhanced variants of the {\em Decorator} pattern is clearly defined and brings important advantages over the classical one.
The fact that through the {\em Decorator} pattern only linear combinations of features are allowed, could be seen as a disadvantage, but using these new variants, this 
is hidden: all the new features become visible. 
Other,  classical examples (e.g  graphical windows) or more complex examples (e.g. features based collection definitions) could be given.


{\bf Acknowledgements}
The complete code of the examples given in the text could be found at \url{https://www.cs.ubbcluj.ro/~vniculescu/Decorator/}

\bibliographystyle{unsrtnat}
\bibliography{mix2}

\begin{thebibliography}{19}
\providecommand{\natexlab}[1]{#1}
\providecommand{\url}[1]{\texttt{#1}}
\expandafter\ifx\csname urlstyle\endcsname\relax
  \providecommand{\doi}[1]{doi: #1}\else
  \providecommand{\doi}{doi: \begingroup \urlstyle{rm}\Url}\fi

\bibitem[Gamma et~al.(1994)Gamma, Helm, Johnson, and Vlissides]{gof}
Erich Gamma, Richard Helm, Ralph Johnson, and John Vlissides.
\newblock \emph{Design Patterns: Elements of Reusable Object Oriented
  Software}.
\newblock Addison Wesley, Oct 1994.

\bibitem[Shalloway and Trott(2004)]{ST04}
Alan Shalloway and James~R. Trott.
\newblock \emph{Design Patterns Explained: A New Perspective on Object Oriented
  Design, 2nd Edition}.
\newblock Addison Wesley, 2004.

\bibitem[B{\"u}chi and Weck(2000)]{generic_wrappers}
Martin B{\"u}chi and Wolfgang Weck.
\newblock Generic wrappers.
\newblock In Elisa Bertino, editor, \emph{ECOOP 2000 --- Object-Oriented
  Programming}, pages 201--225, Berlin, Heidelberg, 2000. Springer Berlin
  Heidelberg.

\bibitem[Kniesel et~al.(2004)Kniesel, Rho, and
  Hanenberg]{Kniesel2004EvolvablePI}
G{\"u}nter Kniesel, Tobias Rho, and Stefan Hanenberg.
\newblock Evolvable pattern implementations need generic aspects.
\newblock In \emph{RAM-SE}, 2004.

\bibitem[Sabane et~al.(2016)Sabane, Gu{\'e}h{\'e}neuc, Arnaoudova, and
  Antoniol]{Sabane2016FragileBP}
Aminata Sabane, Yann-Ga{\"e}l Gu{\'e}h{\'e}neuc, Venera Arnaoudova, and
  Giuliano Antoniol.
\newblock Fragile base-class problem, problem?
\newblock \emph{Empirical Software Engineering}, 22:\penalty0 2612--2657, 2016.

\bibitem[Bloch(2017)]{effective_java}
Joshua Bloch.
\newblock \emph{Effective Java, 3rd Edition}.
\newblock Addison-Wesley Professional, 2017.

\bibitem[Niculescu(2015)]{Nic15}
Virginia Niculescu.
\newblock Mixdecorator: An enhanced version of decorator pattern.
\newblock In \emph{Proceedings of the 20th European Conference on 
  Languages of Programs}, EuroPLoP '15, pages 36:1--36:12, New York, USA, 2015.
  ACM.

\bibitem[Oracle(2018)]{DefaultMethods}
Oracle.
\newblock {Java SE} 8: Implementing default methods in interfaces.
\newblock [online: \url{
  http://www.oracle.com/webfolder/technetwork/tutorials/obe/java/JavaSE8DefaultMethods/JavaSE8DefaultMethods.html}],
  2018.
\newblock Accessed: {July 20, 2020}.

\bibitem[Microsoft(2020)]{CSharp8}
Microsoft.
\newblock What's new in {C}\# 8.0.
\newblock [online: \url{
  https://docs.microsoft.com/en-us/dotnet/csharp/whats-new/csharp-8}], 2020.
\newblock Accessed: {July 20, 2020}.

\bibitem[Kotlin(2020)]{kotlin}
Kotlin.
\newblock Kotlin - extension methods.
\newblock [online: \url{
  https://kotlinlang.org/docs/reference/extensions.html}], 2020.
\newblock Accessed: {August 07, 2020}.

\bibitem[Bono et~al.(2014)Bono, Mensa, and Naddeo]{JavaTraits}
Viviana Bono, Enrico Mensa, and Marco Naddeo.
\newblock Trait-oriented programming in java 8.
\newblock In \emph{Proceedings of the 2014 International Conference on
  Principles and Practices of Programming on the Java Platform: Virtual
  Machines, Languages, and Tools}, PPPJ '14, pages 181--186, New York, NY, USA,
  2014. ACM.

\bibitem[{Microsoft}(2015)]{CS}
{Microsoft}.
\newblock Extension methods ({C}\# programming guide).
\newblock [online:
  \url{https://msdn.microsoft.com/en-us/library/bb383977.aspx}], 2015.
\newblock Accessed: {February 14, 2020}.

\bibitem[Alexandrescu(2001)]{Alexandrescu2001ModernCD}
Andrei Alexandrescu.
\newblock \emph{Modern C++ design: generic programming and design patterns
  applied}.
\newblock Addison-Wesley, 2001.

\bibitem[Abrahams and Gurtovoy(2003)]{metaprogramming04}
David Abrahams and Aleksey Gurtovoy.
\newblock \emph{C++ Template Metaprogramming: Concepts, Tools, and Techniques
  from Boost and Beyond}.
\newblock Addison-Wesley, 2003.

\bibitem[Smaragdakis and Batory(2000)]{C++Mixins}
Y.~Smaragdakis and D.~Batory.
\newblock Mixin-based programming in c++.
\newblock In \emph{Proceedings of the Second International Symposium on
  Generative and Component-Based Software Engineering}, 2000.

\bibitem[Bracha and Cook(1990)]{mixin}
Gilad Bracha and William Cook.
\newblock Mixin-based inheritance.
\newblock In \emph{Proceedings of the European Conference on Object-oriented
  Programming on Object-oriented Programming Systems, Languages, and
  Applications}, OOPSLA/ECOOP '90, pages 303--311, New York, NY, USA, 1990.
  ACM.

\bibitem[Niculescu(2020)]{Nic20}
Virginia Niculescu.
\newblock Efficient decorator pattern variants through c++ policies.
\newblock In \emph{Proceedings of the 15th International Conference on
  Evaluation of Novel Approaches to Software Engineering - Volume 1: ENASE,},
  pages 281--288. INSTICC, SciTePress, 2020.

\bibitem[Sch{\"a}rli et~al.(2002)Sch{\"a}rli, Ducasse, Nierstrasz, and
  Black]{Schrli2002TraitsCU}
Nathanael Sch{\"a}rli, St{\'e}phane Ducasse, Oscar Nierstrasz, and Andrew~P.
  Black.
\newblock Traits: Composable units of behaviour.
\newblock In \emph{ECOOP}, 2002.

\bibitem[Scala(2015)]{ScalaTraits}
Scala.
\newblock Scala documentation: Traits.
\newblock [online: \url {http://docs.scala-lang.org/tour/traits.html}], 2015.
\newblock Accessed: {July 20, 2020}.

\end{thebibliography}

\end{document}